\documentclass[aps,preprint]{revtex4-1}

\usepackage{graphicx}
\usepackage{amsmath}
\newcommand{\R}{\mathbf{r}}
\newcommand{\Y}{\mathbf{y}}

\begin{document}

\title{Locality of contacts determines the subdiffusion exponents in polymeric models of chromatin}
\author{Edoardo Marchi}
\affiliation{Department of Physics, Universit\'a degli Studi di Milano and INFN, via Celoria 16, 20133 Milano, Italy}
\author{Yinxiu Zhan}
\affiliation{Department of Experimental Oncology, European Institute of Oncology IRCCS, via Adamello 16, 20139 Milano, Italy}
\author{Guido Tiana}
\email{guido.tiana@unimi.it}
\affiliation{Department of Physics and Center for Complexity and Biosystems, Universit\'a degli Studi di Milano and INFN, via Celoria 16, 20133 Milano, Italy}
\date{\today}

\begin{abstract}
Loop extrusion by motor proteins mediates the attractive interactions in chromatin on the length scale of megabases, providing the polymer with a well-defined structure and at the same time determining its dynamics. The mean square displacement of chromatin loci varies from a Rouse--like scaling to a more constrained subdiffusion, depending on cell type, genomic region and time scale.
With a simple polymeric model, we show that such a Rouse--like dynamics occurs when the parameters of the model are chosen so that contacts are local along the chain, while in presence of non--local contacts, we observe subdiffusion at short time scales with exponents smaller than 0.5. Such exponents are independent of the detailed choice of the parameters and build a master curve that depends only on the mean locality of the resulting contacts. We compare the loop-extrusion model with a polymeric model with static links, showing that also in this case only the presence of non--local contacts can produce low--exponent subdiffusion. We interpret these results in terms of a simple analytical model.
\end{abstract}

\maketitle

\bibliographystyle{apsrev}

\section{Introduction}

During interphase, chromosomes assume a globular phase in the cellular nucleus characterized by well--defined, although rather mobile domains. Domains at the scale of megabases \cite{Nora2012,Dixon2012,Giorgetti2014,Tiana2016} are called 'topological associating domains' (TADs) and are quite relevant for the control of gene activity \cite{Zhan2017}. 

The mutual attraction between chromosomal regions at this scale is regarded to be mediated by a loop--extrusion mechanism \cite{Fudenberg2016a}. This is based on the action of cohesin, a ring--like protein that binds randomly to the chromatin fiber and extrudes actively the two arms of the fiber, consuming energy. When cohesin meets CCCTC-binding factor (CTCF), a protein bound at specific sites of the DNA, it stops extruding until it is either released from the fiber or it succeeds in overcoming CTCF. Thus, cohesin mediates an effective interaction that is stronger between sites containing CTCF \cite{Crippa2020}. In fact, the removal of cohesin results in the disappearance of TADs \cite{Schwarzer2017}.

Such a mechanism is intrinsically out of equilibrium and it is quite different from that of stabilizing chromatin at larger scales. For example, at the scale of hundreds of megabases, where one can identify 'compartments', the mutual attraction between chromosomal regions is thought to be mediated by DNA--binding proteins, like HP1 \cite{Canzio2011,Zenk2021}, that dimerize, thus stabilizing loops in a framework of equilibrium thermodynamics.

Recent experiments have shown that chromosomal regions undergo anomalous diffusion with different scaling exponents, depending on cell type, genomic region, and time scale.  In the case of mouse embryonic stem cells (mESCs) the scaling exponent is $\approx 0.55$ over the time scale of minutes  \cite{Mach2022,Gabriele23}, where CTCF--mediated contacts assemble and disassemble. This is the same exponent of a Rouse chain with excluded volume \cite{Tamm2017}, which is the least--structured possible polymer. Surprisingly, this scaling exponent does not change if cohesin is depleted. This dynamical freedom is unexpected since cohesin strongly constrains the equilibrium structure of TADs. On the other hand, in HeLa and HT-1080 cells, the scaling exponents decrease to 0.45 and 0.38, respectively \cite{Nozaki2023}. Moreover, in HeLa cells different exponents are observed for euchromatin--rich and heterochromatin--rich regions (0.44 and 0.39, respectively \cite{Shinkai2016}).

We present an analysis of the dynamics of simple polymeric models of chromatin interacting with a loop--extrusion mechanism. The goal is to study the conditions under which the chain can move more freely, in a way that is indistinguishable from that of a Rouse--like chain, and if there are conditions when the subdiffusion is not Rouse--like. We show that the key quantity that controls the diffusion character of the beads of the chain is the average linear separation of the beads that make the contacts.

The idea that interactions between non--consecutive beads in a Rouse model can affect its diffusion properties was introduced in ref. \cite{Amitai2013}, in which it is shown that modifying the eigenvalues that define the rates of the normal modes of the chain by a specific potential leads to non--standard exponents of the mean square displacement of the beads. Moreover, dynamical loops between close beads, which can be formed and disrupted stochastically, have been shown to produce anomalously--low subdiffusion exponents in a polymeric model of chromatin at the scale of compartments \cite{Bohn2010}.

We first present the results of a minimal polymeric model in which the motion of the chain in 3D space is coupled with the motion of extruders along the chain, studying the mean square displacement of the beads (Sect. \ref{sect:le}). Then we study the effect of CTCF bound to the chain, modeled as obstacles that block the 1D motion of the extruders (Sect. \ref{sect:ctcf}). The results obtained simulating the extruders are compared with a simpler polymeric model with static links (Sect. \ref{sect:links}), for which an analytical argument explaining the low subdiffusion exponents can be worked out (Sect. \ref{sect:analytical}).

\section{Subdiffusion under loop extrusion} \label{sect:le}

We simulated the dynamics of a polymeric model made of beads connected by harmonic springs and subject to hard--core repulsion. Extruders are described by other harmonic springs connecting pairs of beads whose identity depends on time. The dynamics of the system is described by a  Langevin equation
\begin{multline}
    m\frac{d\mathbf{v}_n}{dt}=-\gamma \mathbf{v}_n-\nabla_n \left[\frac{k}{2}(|\R_{n+1}-\R_n|-a)^2\right]-\\ -\nabla_n \left[\frac{k}{2}(|\R_{n-
1}-\R_n|-a)^2\right]-\sum_{m=1}^N \nabla_n U_{LJ}(|\R_n-\R_m|)+\\+\frac{k'}{2}\sum_{m\in L_n(t)}\nabla_n(|\R_n-\R_m|-a)^2+\mathbf{\eta}(t),
\end{multline}
where $r_n$ is the position vector of the $n$th bead, $\nabla_n$ is the gradient operator for the $n$th particle in Cartesian coordinates, $k$ is the harmonic constant of the polymer, $\gamma$ is the friction coefficient, $U_{LJ}(r)$ is a truncated Lennard--Jones potential $\epsilon[(r_0/r)^{12}-(r_0/r)^{6}]$ for $r<2^{1/6}r_0$ and zero otherwise, $k'$ the harmonic constant of the extruder, $L_n(t)$ is the time--dependent set of beads connected by an extruder to bead $n$ and $\eta(t)$ is a Gaussian noise with 
$\overline{\eta_n(t)}=0$ and $\overline{\eta_i(t)\eta_j(t')}=6\gamma T\delta_{ij}\delta(t-t')$.

The dynamics of the extruders is modeled as a Markov process. An extruder can be loaded with rate $k_{l}$ on the polymer to connect sites $i$ and $i+1$, where $i$ is a site chosen at random with uniform probability. A bound extruder can leave the polymer with rate $k_u$. An extruder bound to sites $i$ and $j$ ($i<j$) can step  to sites $i-1$ and $j+1$ with rate $k_s$ if these are free; extruders cannot overcome each other. This Markov process is simulated with the Gillespie algorithm \cite{Gillespie1976} along with the Langevin simulation.

The resting distance $a$ of the beads is set to 1 in the simulations and is chosen to represent a segment of 8 kbp, corresponding to approximately 70 nm; we also chose $r_0=1$. We set the temperature of the system to $T=10^4$, which is the order of magnitude of biological temperature if expressed in J/mol (thus the Boltzmann's constant is set to 1); the energy scale $\epsilon$ of the repulsive Lennard--Jones potential is also set to 1. The friction coefficient is $\gamma=10^5$, in such a way that the mean time of displacement of a bead $\tau_b\equiv a^2\gamma/6T\sim 1$, interpreted in seconds, gives the experimental order of magnitude \cite{Mach2022}. The masses are set to $m=10^3$, so that the damping time is $\tau_D\equiv m/\gamma\sim 10^{-2}\ll 1$; at the timescales longer than $\tau_D$, in which we are interested, the motion is over-damped and the results are independent on $m$. The unloading rate of cohesin is $k_u=10^{-2}$ (thus, in s$^{-1}$), as measured from fluorescence experiments \cite{Holzmann2019}; consequently $\tau_D\ll \tau_b\ll k_u^{-1}$. The other two rates $k_l$ and $k_s$ are varied, exploring their effect in the dynamics of the model. The motion of the polymer of $N=10^3$ beads is simulated with a velocity--Verlet algorithm with time step $\Delta t=2.5\cdot 10^{-3}$ s. Simulations are performed with a modified version of the Lammps code \cite{Thompson22,Mach2022}.

For each choice of the parameters, we first equilibrated the system for $10^5$ s. Then, the mean square displacement (MSD) for five of the central beads $n$ of the polymer ($100<n<900$) selected at random is calculated from simulations of $10^6$ s. Assuming that the initial equilibration is long enough to make the MSD homogeneous in time, we used the formula
\begin{equation}
    \text{MSD}(t)=\frac{1}{N_t}\sum_{k=0}^{N_t} \left[ \R(k\cdot\tau_p+t)-\R(k\cdot\tau_p) \right]^2,
\end{equation}
where we recorded $N_t=10^4$ conformations, one every $\tau_p=10^2$ s, also averaging over the 5 beads. Since $\tau_p\gg\tau_D$, we expect the system to behave as over-damped and inertia not to play any role.
 
\begin{figure}
    \centering
    \includegraphics[width=\linewidth]{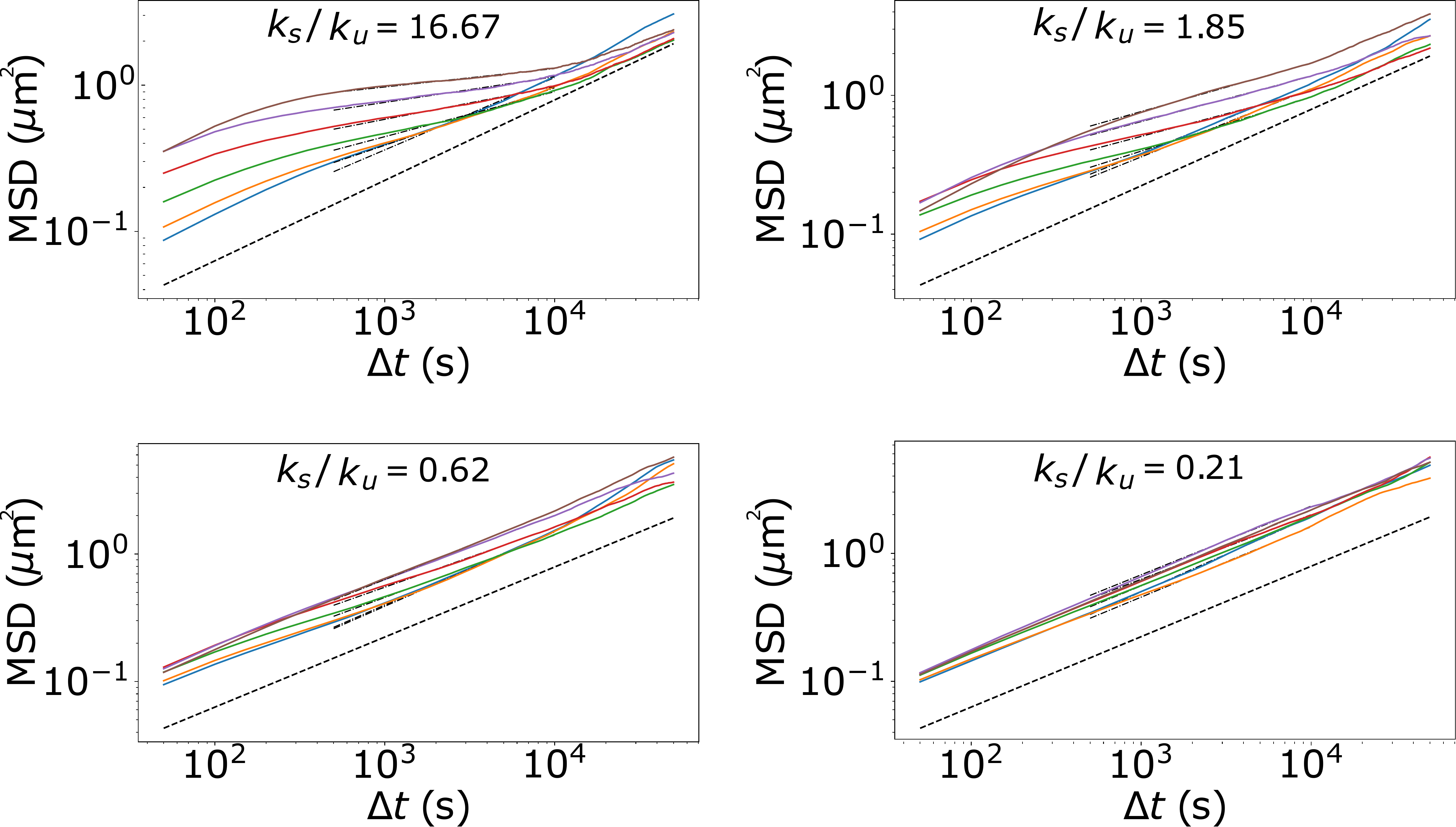}
    \caption{Examples of MSD calculated keeping fixed $k_u=0.1$s$^{-1}$ and varying the other parameters, $k_s/k_u$=[0.21, 0.62, 1.85, 16.67] and $k_l/k_u$=[0.1 (brown, upper curve), 0.31 (violet), 0.93 (red), 2.78 (green), 8.33 (orange), 25.00 (cyan, lower curve)].  The dashed line corresponds to the exponent $0.55$, while the dot--dashed lines are the power--law fits.}
    \label{fig:extrAll}
\end{figure}

The shape of the MSD varies according to the choice of $k_l$ and $k_s$ (some examples are in Fig. \ref{fig:extrAll}). For small $k_s$, all the choices of $k_l$ give a power law $t^\alpha$ with an exponent $\alpha$ compatible with the value $0.55$ that one expects for a Rouse chain with excluded volume  \cite{Tamm2017}. When $k_s$ is increased, the exponent at small times is not monotonic (cf. ref. \cite{Mach2022}) and a new regime appears on the intermediate time scale ($\sim 10^4$ s) with $\alpha<1/2$, corresponding to a more constrained dynamics; we shall focus our attention on this time scale. The value of $\alpha$ here decreases with $k_l$.

\begin{figure*}
    \centering
    \includegraphics[width=\linewidth]{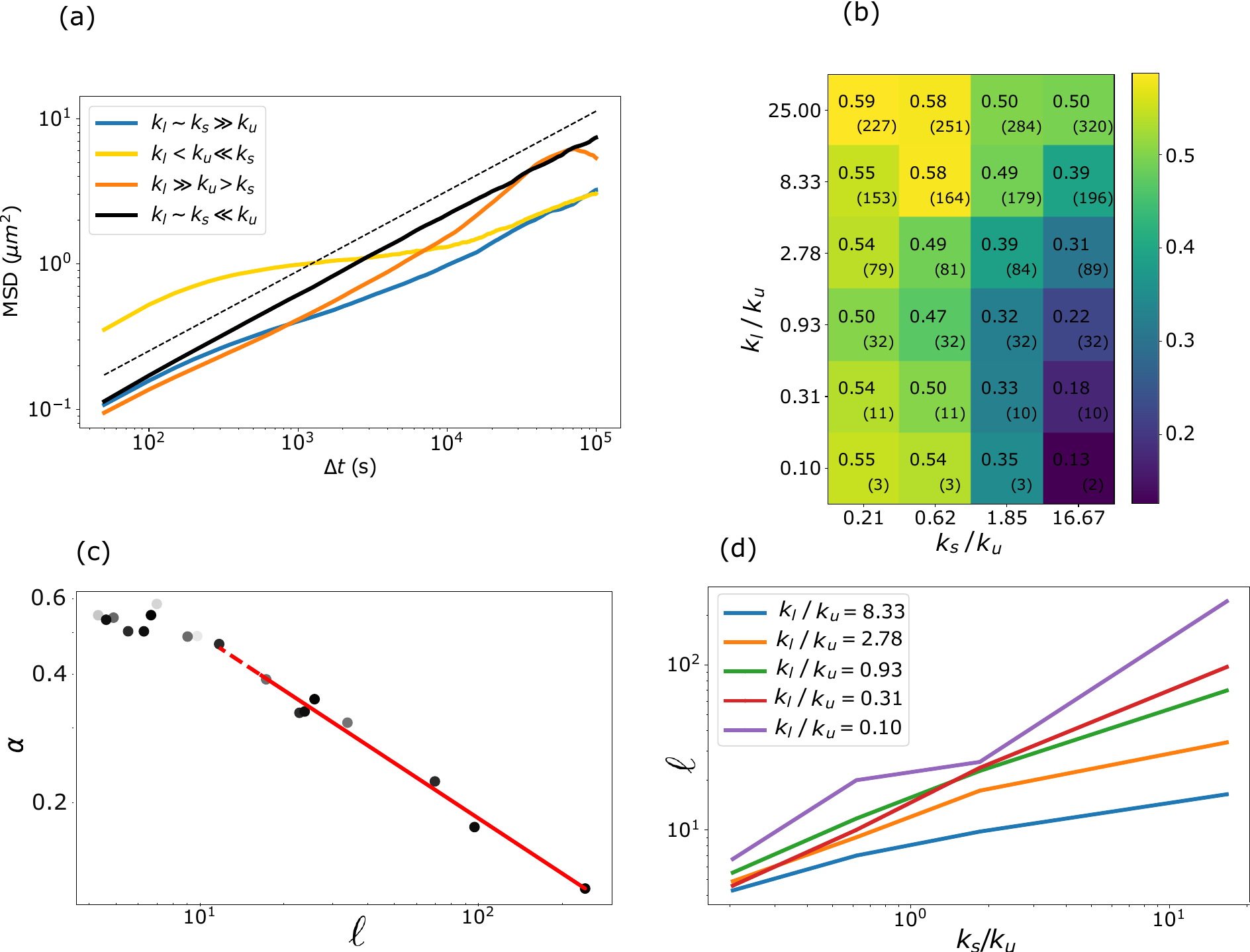}
    \caption{Dynamics of polymer model with loop-extrusion. \textbf{(a)} Examples of mean squared displacement of single beads as a function of time $t$.  \textbf{(b)} Heat--behaviormap of the scaling exponent $\alpha$ for different values of cohesin loading rates and extruders rates. Blue (darker background) indicates sub-Rouse regimes while yellow (lighter background) indicates a super--Rouse regime; within parentheses the average number of active extruders. \textbf{(c)} The scaling exponent $\alpha$ as a function of the mean loop size; darker points indicate a smaller number of bound extruders. \textbf{(d)} Mean loop size, computed as the average genomic distance between linked beads, for the different ratios $k_l/k_u$ ranging from $0.10$ (blue, lower curve) to $8.33$ (purple, upper curve).  }
    \label{fig:extrSum}
\end{figure*}

More specifically, when both the extrusion rate $k_s$ and the loading rate $k_l$ are small, the dynamics is hardly distinguishable from that of a Rouse polymer with excluded volume; higher values of $k_s$ produce a sub-Rouse dynamics at intermediate times  whose exponent depends on the loading rate. Particularly low exponents are obtained when $k_s \gg k_l$, that is when few extruders are moving at high velocity on the chain, forming very large loops (Fig. \ref{fig:extrSum}a). 

At large times one expects to observe only the diffusion ($\alpha=1$) of the center of mass of the polymer. The Rouse time, corresponding to the onset of this regime, is expected to be of the order of $N^2\tau_b\sim 10^6$ s, and thus this regime is not observed in our simulations. 

Realistic parameters of the model can be determined from experiments for a chromatin segment of $N=1000$ sites of $8$ kbp each and are $k_s=10^{-1}$ s$^{-1}$ \cite{Davidson2019} and $k_l=1$ s$^{-1}$ \cite{Mach2022}. This corresponds to the regime in which the motion is Rouse--like, not far from the border of the phase diagram (Fig. \ref{fig:extrSum}b) with the regime characterized by $\alpha<1/2$ that can be reached increasing $k_s$. 

Interestingly, the complexity of the behaviors observed for different choices of $k_s$ and $k_l$ can be easily rationalized in terms of the mean length $\ell\equiv\overline{|i-j|}$ of the loops of the polymer, generated by cohesin binding at sites $i$ and $j$. The exponent $\alpha$ of the intermediate time region depends on the mean loop length independently of the specific value of the rates (Fig. \ref{fig:extrSum}c). The master curve that describes the intermediate---time exponents can be fitted from
\begin{equation}
    \alpha = \ell^{-0.43}
    \label{eq:master}
\end{equation}
for $\ell\gtrsim 10$ and the standard Rouse--like (with excluded volume) value 0.55 for smaller $\ell$.

An observation that is not straightforward is that $\ell$ decreases when increasing the loading rate $k_l$ (Fig. \ref{fig:extrSum}d). This is a consequence of the fact that extruders cannot overcome each other; the result of crowding the polymer with extruders by increasing $k_l$ is not to increase the chance that they can go further along the chain, making $\ell$ larger, but is to increase the probability that they get stuck, decreasing $\ell$.

\section{The effect of obstacles} \label{sect:ctcf}

The extrusion of chromatin by cohesin is stopped by the CTCF protein that binds to specific loci and acts as an obstacle, stabilizing the interaction between the corresponding pair of loci \cite{Fudenberg2016a}.

\begin{figure}
    \centering
    \includegraphics[width=\linewidth]{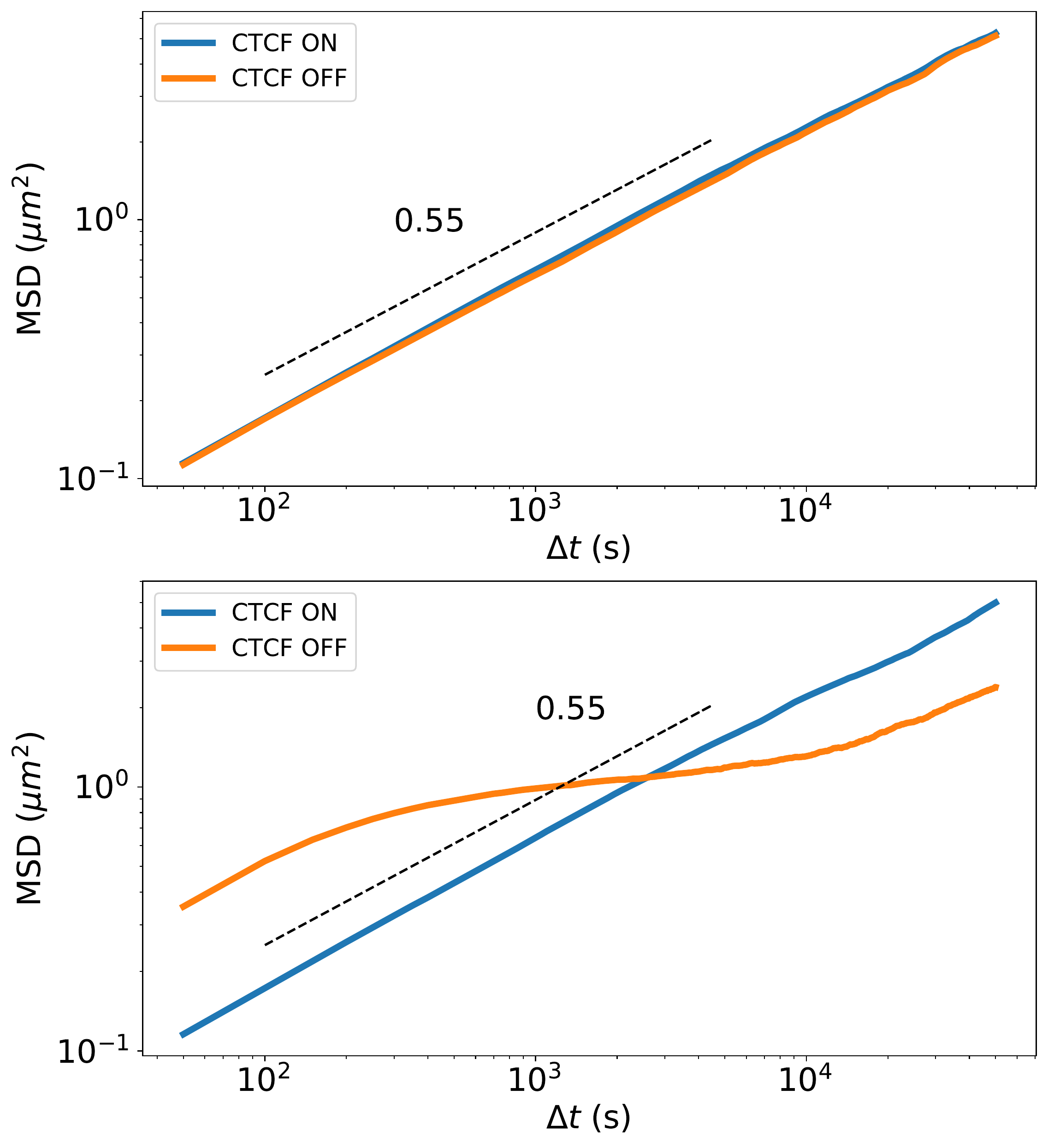}
    \caption{Comparison of dynamics obtained with (blue, darker curves) and without (orange, lighter curves) CTCF. \textbf{(a)} When the extruders speed $k_s$ is low ($k_s/k_u = 0.21$, $k_l/k_u = 0.1$), the loops usually don't reach the barriers; in this case, the dynamics is not affected by CTCF and remains Rouse-like. \textbf{(b)} When $k_s\gg k_u$ ($k_s/k_u = 16.67$, $k_l/k_u = 0.1$)  the loops grow rapidly and often get stuck on the barriers. The mean loop size $\ell$ is thus limited by the mean distance between CTCF sites (using a density $0.1$ of CTCF sites the mean distance between barriers and maximum loop size is $\sim N/100 \ll N$). }
    \label{fig:ctcf}
\end{figure}

We repeated the simulations using a model similar to that of Sect. \ref{sect:le} but introducing CTCF in 10\% of sites chosen at random with uniform distribution. When an extruder reaches a CTCF molecule it cannot cross it but can only unbind from the fiber (in this simple model we do not specify the direction of CTCF, which in chromosomes is asymmetric). For the choice of the parameters that in the absence of CTCF gives a Rouse--like behavior (i.e., exponent 0.55), the presence of CTCF has no effect (upper panel of Fig. \ref{fig:ctcf}). On the contrary, for the parameters that gives a different kind of subdiffusion at different time scales, CTCF makes the system Rouse--like (lower panel of Fig. \ref{fig:ctcf}).

This behavior agrees with the idea that the non--Rouse behavior is due to the presence of long--range contacts; the presence of CTCF blocks the extrusion and reduces the range of the contacts to linear distances which are of the order of the mean spacing between consecutive CTCF molecules, in this case, $\approx 10$ beads.

\section{A simpler model with quenched links} \label{sect:links}

To gain a deeper understanding of the effect of long--range contacts on the dynamics of the polymer, we have studied the MSD of a simpler model in which, instead of being mediated by moving extruders, random pairs of beads are linked together by quenched harmonic springs (with the same constant as the springs that maintain the topology of the chain). In this way, we can disentangle the time--dependent effect of the kinetics of the extruders from that of the long--range contacts they can make when we interpret the results of Sect. \ref{sect:le} . We tested different numbers of links and different probability distributions from which to extract the linked pairs, defined as a function of $\ell$. We also run some simulations with no excluded volume to check the effects of hard-core repulsion on the dynamics. 

\begin{figure*}
    \centering
    \includegraphics[width=\linewidth]{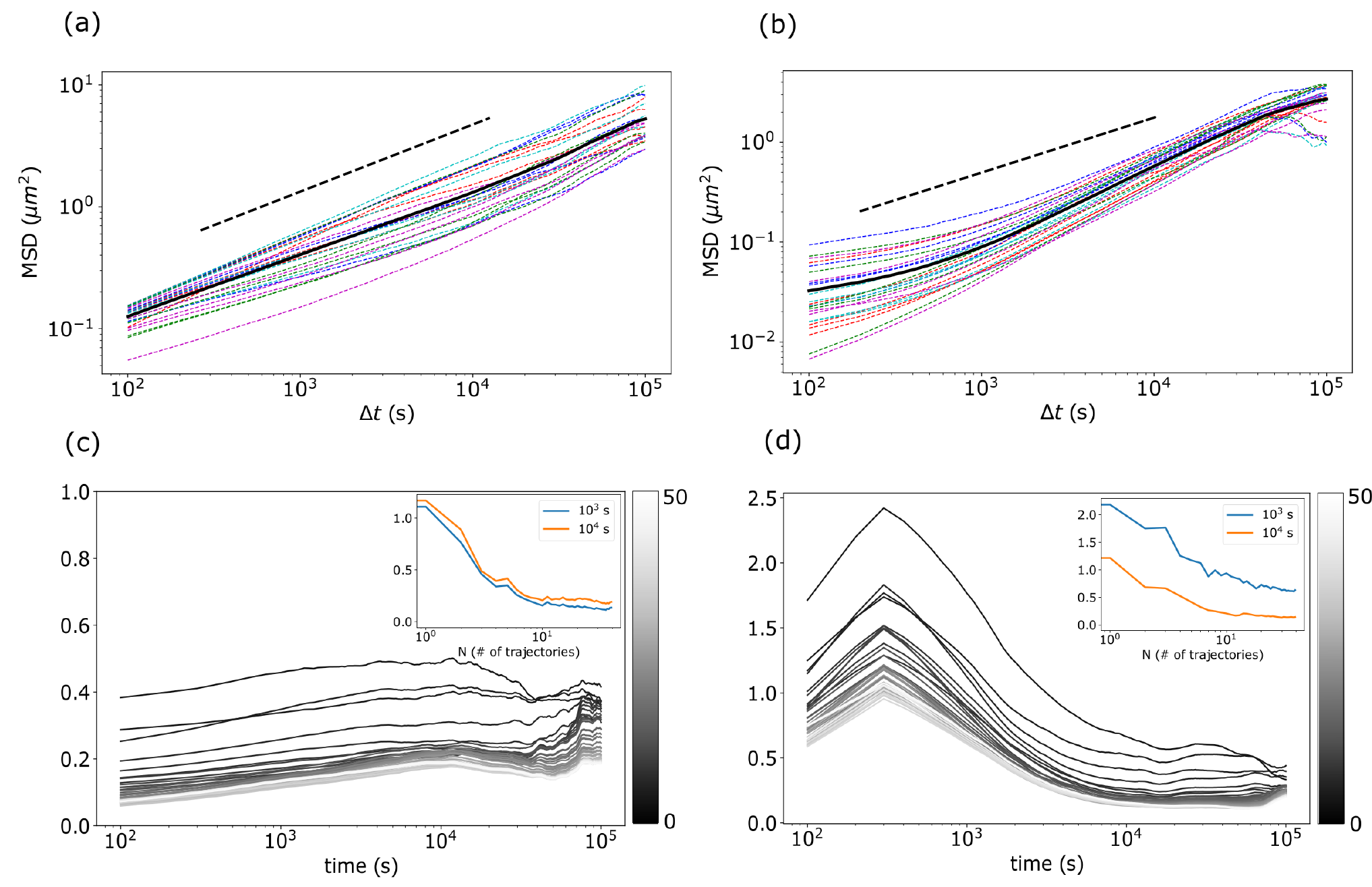}
    \caption{Spread of trajectories and the self-averaging property. (a) The realizations of the MSD of a chain of 1000 beads with $\lambda=10$ fixed links extracted from a uniform distribution; lines with the same color (gray level) come from different beads of the same realization of links. The full black line is the average of all the trajectories. The dashed line is the reference with a slope 0.55. (b) The same with $l=100$ links. (c) The self--averageness parameter $\rho(t)$ for a different number $n$ of instances when $\lambda=10$. The inset shows the curves of the relative variance as a function of the number of trajectories $N$, at $t=10^3$s (blue, darker line) and $t=10^4$s (orange, lighter curve). (d) The same with $\lambda=100$ links. }
    \label{fig:linksAverage}
\end{figure*}

We have first simulated the dynamics of a chain of $N=1000$ beads interacting with $\lambda=10$ or $\lambda=100$ links extracted by a uniform distribution. For both cases, we generated 5 independent realizations of the links and studied the MSD of 10 beads randomly extracted from the central portion of the chain ($100<i<900$). For $\lambda=10$, some of the curves of the MSD$(t)$ display a power law with exponent $\approx 0.55$, but others are slightly bent with both kinds of concavity (Fig. \ref{fig:linksAverage}a). The average curve anyway displays the single--exponent power law typical of Rouse--like chain with excluded volume.

If we increase the number of links (Fig. \ref{fig:linksAverage}b) all the realizations of the dynamics display a multi--exponent subdiffusion at small time scales, that eventually converge to $\alpha\approx 0.55$, although the instantaneous diffusion coefficient (i.e., the vertical offset) is dependent on the specific realization. The average MSD maintains in this case the two--exponent behavior of the single realizations. 

The interpretation of the average MSD curve as typical of the different realizations is correct only if the MSD is self--averaging \cite{Brout1959}. To assess if this is the case, we have studied the normalized variance
\begin{equation}
    \rho(t)\equiv \frac{\sum_{k=1}^n \text{MSD}^2_k(t)}{\left[\sum_{k=1}^n \text{MSD}_k(t)\right]^2}-1
\end{equation}
of $n$ instances MSD$_k(t)$ of the dynamics ($1<k\leq n$), as a function of $n$ (Fig. \ref{fig:linksAverage}c and d).

For both choices $\lambda=10$ and $\lambda=100$, the curves of $\rho(t)$ decrease with respect to $n$ (Fig. \ref{fig:linksAverage}c and d). The decrease of $\rho$ is clearer if studied as a function of $n$ at selected times ($t=10^3$s and $t=10^4$s, insets in Figs. \ref{fig:linksAverage}c and \ref{fig:linksAverage}d). This suggests that the MSD is a self--averaging property of the system.

\begin{figure}
    \centering
    \includegraphics[width=\linewidth]{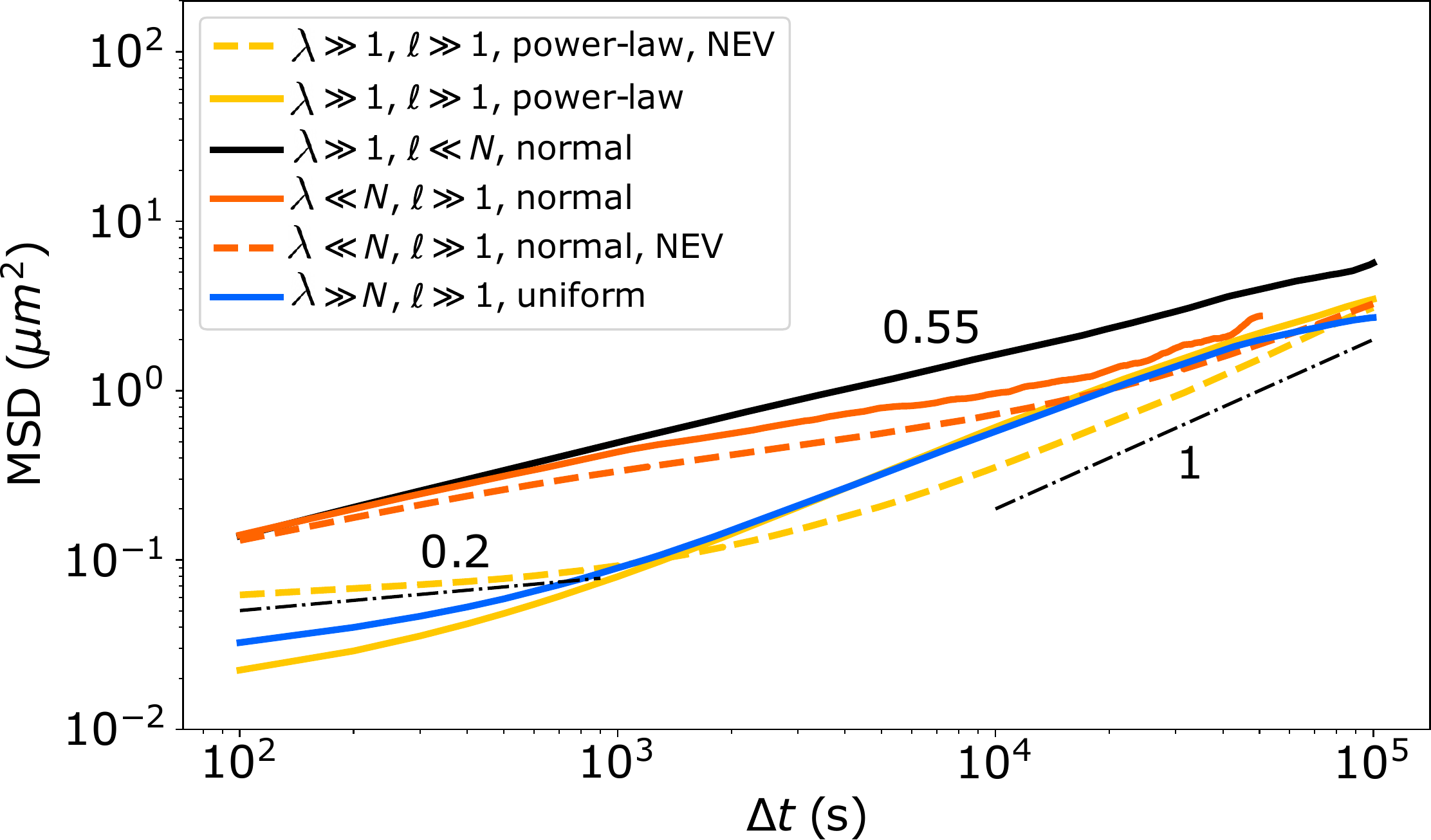}
    \caption{Dynamics of polymer model with static links. Some examples of the mean squared displacement of beads for the polymers with quenched links with different numbers of links $\lambda$ and different links distributions (uniform, power-law, normal). Some of the simulations are performed without excluded volume (NEV, dashed lines).}
    \label{fig:linksExamples}
\end{figure}

The shape of the MSD, averaged over the quenched disorder and on the choice of the bead, depends on the form of the distribution and on the number $l$ of links (Fig. \ref{fig:linksExamples}). For example, if we choose links between beads $i$ and $j$ from a normal distribution with average $d$,
\begin{equation}
    p(i,j)\propto \exp\left[-\frac{(|i-j|-d)^2}{10} \right],
\end{equation}
if $\ell=d\ll N$ the MSD is Rouse--like even in the presence of a large number of links ($\lambda\sim N$). Vice versa, a small number ($\lambda\ll N$) of long--range links ($\ell=d\sim N$) is sufficient to bend the MSD and have a short--time subdiffusion with an exponent smaller than 0.55.

Choosing a power--law distribution
\begin{equation}
 p(i,j)\propto \frac{1}{|i-j|^{\beta}}   
\end{equation}
with $\beta=0.8$ or a uniform distribution ($\beta=0$) which yield large loops $\ell\gg 1$, the dynamics on the intermediate--time regime is sub-Rouse with exponents as low as $\alpha \approx 0.2$. For long times the MSD approaches a linear scaling $\alpha = 1$, suggesting that one of the effects of links between beads is to decrease Rouse time, anticipating the effect of the diffusion of the center of mass. Finally, we observe that by removing the interactions that cause excluded volume, the range of times associated with low exponents is shifted forward.

\begin{figure}
    \centering
    \includegraphics[scale = 0.37]{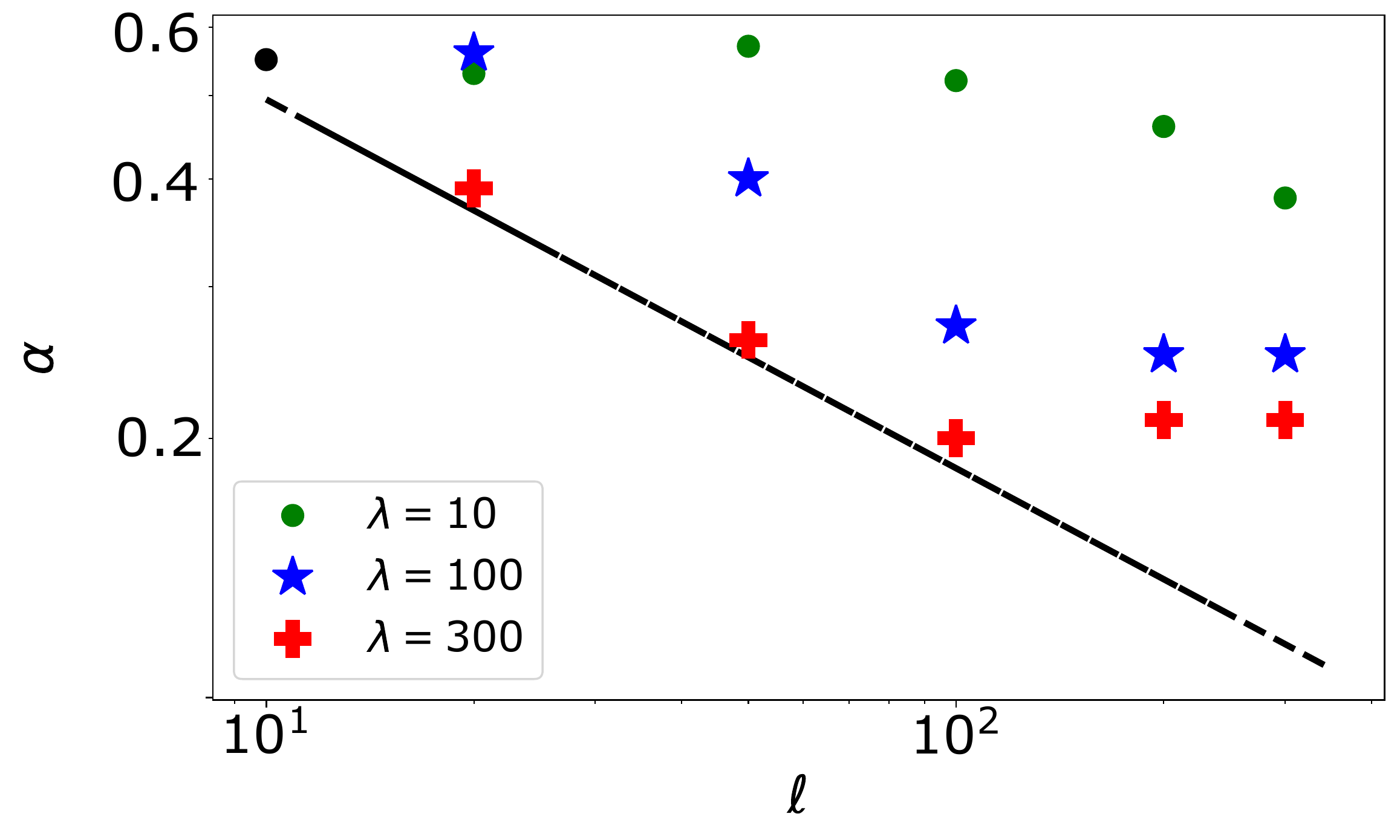}
    \caption{Relationship between $\alpha$ and $\ell$ with static links. The  scaling exponent $\alpha$ as a function of the mean loop size $\ell$ for 3 different choices of the number of links. The black line is the fitting curve obtained from Fig. \protect\ref{fig:extrSum} with loop-extrusion. }
    \label{fig:exponentsLinks}
\end{figure}

The exponents $\alpha$ of the MSD for the small--time regime of the chain with fixed links depend not only on the average linear distance $\ell$ between the linked beads but also on the number $\lambda$ of links (red points in Fig. \ref{fig:exponentsLinks}). When the number of links is large, the exponents coincide with those obtained from loop extrusion. Only for very large  $\ell$ the exponent saturates, because the number of pairs of beads that can be linked decreases with $\ell$ decreasing the (effective) number of independent pairs.

When the number of pairs $\lambda$ is smaller (blue and green points in Fig. \ref{fig:exponentsLinks}), $\alpha$ seem to still follow a power--law similar to that at large $\lambda$ but leaving the Rouse--like exponent $0.55$ at larger values of $\ell$.

\section{A simple analytical model} \label{sect:analytical}

The polymer with quenched links allows one to develop a minimal analytical model to investigate the role of long--range contacts. The goal is to show that links that are local along the chain affect only the diffusion coefficient, but not the subdiffusion exponent, while highly nonlocal contacts can change the exponent.
Assume that a Rouse chain \cite{Rouse1953APolymers} of length $2N$ is linked by harmonic springs of constant $k$. Moreover, each bead $i$ interacts with beads $i+m$ with a harmonic spring of constant $k'$. Let's assume that this interaction occurs with a random bead, so that $m$ is a random variable. The Brownian dynamics of the chain is controlled by the Langevin equation
\begin{equation}
\frac{d\R_n}{dt}=\frac{k}{\gamma}(\R_{n+1}+\R_{n-1}-2r_n)-\frac{k'}{\gamma} (\R_n-\R_{n+m})+\frac{1}{\gamma}\mathbf{\eta}_n(t),
\end{equation}
where $-N\leq n\leq N$, $\gamma$ is the friction constant and $\eta(t)$ is a random noise satisfying $\overline{\eta_n(t)}=0$ and $\overline{\eta_i(t)\eta_j(t')}=6\gamma^2 D\delta_{ij}\delta(t-t')$.
In the limit of large $N$ one can consider $n$ as a continuous variable and rewrite the Langevin equation as
\begin{equation}
\frac{\partial \R(n,t)}{\partial t}=\frac{k}{\gamma}\frac{\partial^2 \R}{\partial n^2}-\frac{k'}{\gamma} [\R(n,t)-\R(n+m,t)]+\frac{1}{\gamma}\mathbf{\eta}(n,t).
\end{equation}
Using the boundary conditions as in ref. \cite{Grosberg}, one can study the system in Fourier space by defining
\begin{equation}
\mathbf{y}_p(t)=\frac{1}{2N}\int_{-N}^N dn\; \R(n,t) \cos(\pi p n/2N),
\end{equation}
with $0\leq p\leq N$, one obtains
\begin{equation}
\frac{d\Y_p(t)}{dt}=-\frac{\pi^2k}{N^2\gamma}p^2\Y_p(t)-\frac{k'}{\gamma} \Y_p(t)[1-\cos(\pi p m/2N)] +\frac{1}{\gamma}\mathbf{\eta}_p(t).
\end{equation}
whose solution is
\begin{equation}
\overline{|\R(x,t)|^2}=\frac{3Dt}{N}+\sum_{p\neq 0}\left|\frac{D\tau_p}{N}\left[1-e^{-t/\tau_p} \right]\right|,
\label{eq:r2}
\end{equation}
which is formally the same as the standard Rouse model \cite{Rouse1953APolymers}, but now with 
\begin{equation}
\tau_p\equiv \frac{N^2\gamma}{\pi k p^2+ N^2 k'[1-\cos(\pi p m/2N)]}.
\label{eq:taup}
\end{equation}
Clearly, the extra links affect $\tau_p$ mainly at small $p$, that is at large length scales, while at large $p$ the Rouse term $\pi k p^2$ dominates at the denominator.
The term with the sum in Eq. (\ref{eq:r2}) can be approximated for large $N$ as proportional to the integral
\begin{equation}
I(t)=\int_0^\infty dp\;\left|\tau_p\left[1-e^{-t/\tau_p} \right]\right|,
\label{eq:int}
\end{equation}
whose integrand increases with $\tau_p$ and thus decreases with $p$.

If contacts are local ($m\ll N$), then the cosine can be approximated as $1-\pi^2 m^2 p^2/4N^2$, the integral (\ref{eq:int}) is still Gaussian and one obtains the standard results for the Rouse chain $I(t)\sim t^{1/2}$, just renormalizing the diffusion coefficient.

If contacts are nonlocal, the argument of the cosine is no longer small and the dependence on time can be different. In this case, we expect that if $m\gg 1$, each bead is linked with other beads whose specific index in not relevant; thus, we will assume that $x\equiv 1-\cos(\pi p/2N \cdot m)$ is a random variable independent on $p$. More specifically, if $m\gg 1$ and $p\leq N$, $\cos(\pi p/2N \cdot m)$ fluctuates so rapidly with respect to $p$ that its evaluation at a given $p$ gives a number with the features of a stochastic variable, similarly to pseudo--random numbers generated by the linear congruential algorithm \cite{Thomson1958ANumbers}.
The integral of Eq. (\ref{eq:int}) for given $x$ is then
\begin{equation}
I_x(t)=\frac{\pi^{1/2}N\gamma}{2(kk'x)^{1/2}}\text{erf}\left(\frac{(k'xt)^{1/2}}{\gamma^{1/2}} \right).
\end{equation}
To calculate the distribution of $x$ given a specific value of $m$, one uses
\begin{align}
    P(x|m)&=P(p|m)\left|\frac{dp}{dx} \right|=\\
    &=\theta\left(\frac{2}{\pi m}\arccos{[1-x}]-1\right)\frac{2N}{\pi m\sqrt{x(2-x)}}, \nonumber
\end{align}
where Heaviside's theta comes from the fact that $p\leq N$. Assuming that $m$ is a random variable as well, with distribution $P(m)=m^{-\beta}$ as usually found in chromatin \cite{Mirny2011}, then
\begin{align}
    P(x)&=\int_0^N dm\; P(x|m) P(m)= \\
    &=\frac{1}{[1-2\arcsin{(1-x)}/\pi]^\beta [x(2-x)]^{1/2}} \sim\frac{1}{x^{(\beta+1)/2}}, \nonumber
\end{align}
where $0\leq x\leq 2$.
The expectation value of $I(t)$ with respect to the distribution $P(x)$ is then
\begin{align}
     &E_x[I_x(t)]=\int_0^2 dx\; P(x) I_x(t) = \\
     &=\text{erf}(\sqrt{2t})+t^{\beta/2} \left[\Gamma(1/2-\beta/2,2t)-\Gamma(1/2-\beta/2 )\right], \nonumber
\end{align}
where the $\Gamma$ are the incomplete and the complete, respectively, Euler's function. For large $t$,
\begin{equation}
    E_x[I_x(t)]\sim t^{\beta/2},
\end{equation}
meaning that the MSD can display non--standard exponents if $\beta\neq 1$.

\section{Discussion and Conclusions}

The dynamics of chromatin, quantified by its MSD, is determined by different factors. On one side, it is a polymer with a consistent degree of structure, which is intimately connected with the control of gene activity. Thus, one can expect strong subdiffusion, caused by the constraints that the polymeric bonds and the other interactions exert on the motion of each portion of the chromosome. On the other hand, loop extrusion is an active process, that can cause different types of dynamics, including superdiffusion. 

Experiments show that chromatin undergoes subdiffusion with variable exponents, which are in all available cases similar or somewhat smaller than that of a Rouse chain with excluded volume \cite{Mach2022,Gabriele23,Nozaki2023,Shinkai2016}. This result is not straightforward because a polymer with complex interactions among its parts and with the nuclear lamina, resulting from an active, energy--consuming process, can in principle display any kind of dynamics, from super--diffusive to non--diffusive.
On the other hand, the observed kind of dynamics leaves to the chain the ability to undergo fast conformational changes following the cellular cycle and as a response to stimuli. It is then interesting to ask whether this is the typical dynamics of a polymer caused by loop extrusion or if evolution had to fine--tune the mechanism to achieve this result.

Simulations of a polymer model with loop extrusion show that two types of behavior are possible, either a Rouse--like dynamics with exponent $\alpha=0.55$ (due to the excluded volume) up to the Rouse time, or a more complex subdiffusion that at intermediate times displays an exponent $\alpha<0.5$. Clearly, the latter case corresponds to much slower dynamics, that according to our estimates occur on the time scale of hours to days, which is the time scale of most cellular processes. 

What determines the type of subdiffusion is only the average length $\ell$ of the loops created by the extruders, within our model and in the range of parameters we studied. In fact, the value of the exponent $\alpha$ for the intermediate time window obtained from the simulations is described by a simple function of $\ell$ independently of the parameters used in the simulations. This function is constant (0.55) up to $\ell\approx 10$ and then decreases as $\alpha\approx\ell^{-0.43}$. This fact suggests that it is only the presence of non--local contacts to slow down the dynamics,  not that of any kind of contact. On the length scale of TADs, chromatin display an architecture rich of local contacts, similar to a Peano curve \cite{Mirny2011}, with a hierarchical structure of contacts \cite{Zhan2017}, and thus its dynamics remain Rouse--like. This is mainly due to the presence of CTCF, which hinders the motion of the extruder, reducing $\ell$. 

The role of non--local contacts can be studied in an easier way by introducing static  links in the polymer. A sensible way to implement this is by linking pairs of monomers of the chain, chosen at random with a probability that depends on their distance, with harmonic springs. The links are kept fixed during each simulation and eventually we studied quenched average of the MSD. We first showed that the MSD is self--averaging, meaning that its average is representative of the MSD of a typical realization of the links. Also this model can display subdiffusion with $\alpha<0.5$ and $\alpha$ depends only on $\ell$. If the number of links is large, the function $\alpha(\ell)$ is the same as that obtained with loop extrusion. This fact suggests that the dynamics of the extruder plays no role in determining the MSD, only the range of the resulting contacts is relevant. 

Decreasing the number of links changes the function $\alpha(\ell)$, extending the region of $\ell$ where $\alpha=0.55$ and increasing overall $\alpha$. With a small number of links, it is more difficult for the quenched links to constrain efficiently all the monomers of the polymer, something that the extruders can do moving along the chain.

The effect of the quenched links can be captured by a minimal analytic model that follows the idea of the Rouse derivation \cite{Rouse1953APolymers}. The effect of the non--local contacts is to modify the relaxation times of the normal modes of the polymer in a way that depends on the degree of non--locality. This can change the exponent of the resulting MSD as a function of time. In the limit of local contacts, the relaxation times re--acquire the $1/p^2$ dependence and $\alpha$ returns to that of standard Rouse subdiffusion.

\acknowledgments{We thank Luca Giorgetti for his help and discussions.}

\bibliographystyle{apsrev4-1}

\end{document}